

%
%


\def\famname{
 \textfont0=\textrm \scriptfont0=\scriptrm
 \scriptscriptfont0=\sscriptrm
 \textfont1=\textmi \scriptfont1=\scriptmi
 \scriptscriptfont1=\sscriptmi
 \textfont2=\textsy \scriptfont2=\scriptsy \scriptscriptfont2=\sscriptsy
 \textfont3=\textex \scriptfont3=\textex \scriptscriptfont3=\textex
 \textfont4=\textbf \scriptfont4=\scriptbf \scriptscriptfont4=\sscriptbf
 \skewchar\textmi='177 \skewchar\scriptmi='177
 \skewchar\sscriptmi='177
 \skewchar\textsy='60 \skewchar\scriptsy='60
 \skewchar\sscriptsy='60
 \def\rm{\fam0 \textrm} \def\bf{\fam4 \textbf}}
\def\sca#1{scaled\magstep#1} \def\scah{scaled\magstephalf} 
\def\twelvepoint{
 \font\textrm=cmr12 \font\scriptrm=cmr8 \font\sscriptrm=cmr6
 \font\textmi=cmmi12 \font\scriptmi=cmmi8 \font\sscriptmi=cmmi6 
 \font\textsy=cmsy10 \sca1 \font\scriptsy=cmsy8
 \font\sscriptsy=cmsy6
 \font\textex=cmex10 \sca1
 \font\textbf=cmbx12 \font\scriptbf=cmbx8 \font\sscriptbf=cmbx6
 \font\it=cmti12
 \font\sectfont=cmbx12 \sca1
 \font\sectmath=cmmib10 \sca2
 \font\sectsymb=cmbsy10 \sca2
 \font\refrm=cmr10 \scah \font\refit=cmti10 \scah
 \font\refbf=cmbx10 \scah
 \def\twelverm{\textrm} \def\twelveit{\it} \def\twelvebf{\textbf}
 \famname \textrm 
 \advance\voffset by .06in \advance\hoffset by .28in
 \normalbaselineskip=17.5pt plus 1pt \baselineskip=\normalbaselineskip
 \parindent=21pt
 \setbox\strutbox=\hbox{\vrule height10.5pt depth4pt width0pt}}


\catcode`@=11

{\catcode`\'=\active \def'{{}^\bgroup\prim@s}}

\def\screwcount{\alloc@0\count\countdef\insc@unt}   
\def\screwdimen{\alloc@1\dimen\dimendef\insc@unt} 
\def\screwbox{\alloc@4\box\chardef\insc@unt}

\catcode`@=12


\overfullrule=0pt			
\vsize=9in \hsize=6in
\lineskip=0pt				
\abovedisplayskip=1.2em plus.3em minus.9em 
\belowdisplayskip=1.2em plus.3em minus.9em	
\abovedisplayshortskip=0em plus.3em	
\belowdisplayshortskip=.7em plus.3em minus.4em	
\parindent=21pt
\setbox\strutbox=\hbox{\vrule height10.5pt depth4pt width0pt}
\def\makefootline{\baselineskip=30pt \line{\the\footline}}
\footline={\ifnum\count0=1 \hfil \else\hss\twelverm\folio\hss \fi}
\pageno=1


\def\put(#1,#2)#3{\screwdimen\unit  \unit=1in
	\vbox to0pt{\kern-#2\unit\hbox{\kern#1\unit
	\vbox{#3}}\vss}\nointerlineskip}


\def\\{\hfil\break}
\def\newpage{\vfill\eject}
\def\center{\leftskip=0pt plus 1fill \rightskip=\leftskip \parindent=0pt
 \def\textindent##1{\par\hangindent21pt\footrm\noindent\hskip21pt
 \llap{##1\enspace}\ignorespaces}\par}
\def\unnarrower{\leftskip=0pt \rightskip=\leftskip}


\def\vol#1 {{\refbf#1} }		 


\def\NP #1 {{\refit Nucl. Phys.} {\refbf B{#1}} }
\def\PL #1 {{\refit Phys. Lett.} {\refbf{#1}} }
\def\PR #1 {{\refit Phys. Rev. Lett.} {\refbf{#1}} }
\def\PRD #1 {{\refit Phys. Rev.} {\refbf D{#1}} }


\hyphenation{pre-print}
\hyphenation{quan-ti-za-tion}

%
%


\def\oonoo#1#2#3{\vbox{\ialign{##\crcr
	\hfil\hfil\hfil{$#3{#1}$}\hfil\crcr\noalign{\kern1pt\nointerlineskip}
	$#3{#2}$\crcr}}}
\def\oon#1#2{\mathchoice{\oonoo{#1}{#2}{\displaystyle}}
	{\oonoo{#1}{#2}{\textstyle}}{\oonoo{#1}{#2}{\scriptstyle}}
	{\oonoo{#1}{#2}{\scriptscriptstyle}}}
\def\dt#1{\oon{\hbox{\bf .}}{#1}}  
\def\ddt#1{\oon{\hbox{\bf .\kern-1pt.}}#1}    
\def\slap#1#2{\setbox0=\hbox{$#1{#2}$}
	#2\kern-\wd0{\hfuzz=1pt\hbox to\wd0{\hfil$#1{/}$\hfil}}}
\def\sla#1{\mathpalette\slap{#1}}                
\def\bop#1{\setbox0=\hbox{$#1M$}\mkern1.5mu
	\lower.02\ht0\vbox{\hrule height0pt depth.06\ht0
	\hbox{\vrule width.06\ht0 height.9\ht0 \kern.9\ht0
	\vrule width.06\ht0}\hrule height.06\ht0}\mkern1.5mu}
\def\bo{{\mathpalette\bop{}}}                        
\def~{\widetilde} 
\mathcode`\*="702A                  
\def\in{\relax\ifmmode\mathchar"3232\else{\refit in\/}\fi} 
\def\half{{\textstyle{1\over{\raise.1ex\hbox{$\scriptstyle{2}$}}}}}

\def\Gamma{\mathchar"0100}
\def\Delta{\mathchar"0101}
\def\Theta{\mathchar"0102}
\def\Lambda{\mathchar"0103}
\def\Xi{\mathchar"0104}
\def\Pi{\mathchar"0105}
\def\Sigma{\mathchar"0106}
\def\Upsilon{\mathchar"0107}
\def\Phi{\mathchar"0108}
\def\Psi{\mathchar"0109}
\def\Omega{\mathchar"010A}

\catcode128=13 \def €{\"A}                 
\catcode129=13 \def {\AA}                 
\catcode130=13 \def '{\c}           	   
\catcode131=13 \def ƒ{\'E}                   
\catcode132=13 \def "{\~N}                   
\catcode133=13 \def …{\"O}                 
\catcode134=13 \def †{\"U}                  
\catcode135=13 \def ‡{\'a}                  
\catcode136=13 \def ˆ{\`a}                   
\catcode137=13 \def ‰{\^a}                 
\catcode138=13 \def Š{\"a}                 
\catcode139=13 \def ‹{\~a}                   
\catcode140=13 \def Œ{\alpha}            
\catcode141=13 \def {\chi}                
\catcode142=13 \def Ž{\'e}                   
\catcode143=13 \def {\`e}                    
\catcode144=13 \def {\^e}                  
\catcode145=13 \def '{\"e}                
\catcode146=13 \def '{\'\i}                 
\catcode147=13 \def "{\`\i}                  
\catcode148=13 \def "{\^\i}                
\catcode149=13 \def •{\"\i}                
\catcode150=13 \def –{\~n}                  
\catcode151=13 \def —{\'o}                 
\catcode152=13 \def ˜{\`o}                  
\catcode153=13 \def ™{\^o}                
\catcode154=13 \def š{\"o}                 
\catcode155=13 \def ›{\~o}                  
\catcode156=13 \def œ{\'u}                  
\catcode157=13 \def {\`u}                  
\catcode158=13 \def ž{\^u}                
\catcode159=13 \def Ÿ{\"u}                
\catcode160=13 \def  {\tau}               
\catcode161=13 \mathchardef ¡="2203     
\catcode162=13 \def ¢{\oplus}           
\catcode163=13 \def £{\relax\ifmmode\to\else\itemize\fi} 
\catcode164=13 \def ¤{\subset}	  
\catcode165=13 \def ¥{\infty}           
\catcode166=13 \def ¦{\mp}                
\catcode167=13 \def §{\sigma}           
\catcode168=13 \def ¨{\rho}               
\catcode169=13 \def ©{\gamma}         
\catcode170=13 \def ª{\leftrightarrow} 
\catcode171=13 \def «{\relax\ifmmode\acute\else\expandafter\'\fi}
\catcode172=13 \def ¬{\relax\ifmmode\expandafter\ddt\else\expandafter\"\fi}
\catcode173=13 \def ­{\equiv}            
\catcode174=13 \def ®{\approx}          
\catcode175=13 \def ¯{\Omega}          
\catcode176=13 \def °{\otimes}          
\catcode177=13 \def ±{\ne}                 
\catcode178=13 \def ²{\le}                   
\catcode179=13 \def ³{\ge}                  
\catcode180=13 \def ´{\upsilon}          
\catcode181=13 \def µ{\mu}                
\catcode182=13 \def ¶{\delta}             
\catcode183=13 \def ·{\epsilon}          
\catcode184=13 \def ¸{\Pi}                  
\catcode185=13 \def ¹{\pi}                  
\catcode186=13 \def º{\beta}               
\catcode187=13 \def »{\partial}           
\catcode188=13 \def ¼{\nobreak\ }       
\catcode189=13 \def ½{\zeta}               
\catcode190=13 \def ¾{\sim}                 
\catcode191=13 \def ¿{\omega}           
\catcode192=13 \def À{\dt}                     
\catcode193=13 \def Á{\gets}                
\catcode194=13 \def Â{\lambda}           
\catcode195=13 \def Ã{\nu}                   
\catcode196=13 \def Ä{\phi}                  
\catcode197=13 \def Å{\xi}                     
\catcode198=13 \def Æ{\psi}                  
\catcode199=13 \def Ç{\int}                    
\catcode200=13 \def È{\oint}                 
\catcode201=13 \def É{\relax\ifmmode\cdot\else\vol\fi}    
\catcode202=13 \def Ê{\relax\ifmmode\,\else\thinspace\fi}
\catcode203=13 \def Ë{\`A}                      
\catcode204=13 \def Ì{\~A}                      
\catcode205=13 \def Í{\~O}                      
\catcode206=13 \def Î{\Theta}              
\catcode207=13 \def Ï{\theta}               
\catcode208=13 \def Ð{\relax\ifmmode\bar\else\expandafter\=\fi}
\catcode209=13 \def Ñ{\overline}             
\catcode210=13 \def Ò{\langle}               
\catcode211=13 \def Ó{\relax\ifmmode\{\else\ital\fi}      
\catcode212=13 \def Ô{\rangle}               
\catcode213=13 \def Õ{\}}                        
\catcode214=13 \def Ö{\sla}                      
\catcode215=13 \def ×{\relax\ifmmode\check\else\expandafter\v\fi}
\catcode216=13 \def Ø{\"y}                     
\catcode217=13 \def Ù{\"Y}  		    
\catcode218=13 \def Ú{\Leftarrow}       
\catcode219=13 \def Û{\Leftrightarrow}       
\catcode220=13 \def Ü{\relax\ifmmode\Rightarrow\else\sect\fi}
\catcode221=13 \def Ý{\sum}                  
\catcode222=13 \def Þ{\prod}                 
\catcode223=13 \def ß{\widehat}              
\catcode224=13 \def à{\pm}                     
\catcode225=13 \def á{\nabla}                
\catcode226=13 \def â{\quad}                 
\catcode227=13 \def ã{\in}               	
\catcode228=13 \def ä{\star}      	      
\catcode229=13 \def å{\sqrt}                   
\catcode230=13 \def æ{\^E}			
\catcode231=13 \def ç{\Upsilon}              
\catcode232=13 \def è{\"E}    	   	 
\catcode233=13 \def é{\`E}               	  
\catcode234=13 \def ê{\Sigma}                
\catcode235=13 \def ë{\Delta}                 
\catcode236=13 \def ì{\Phi}                     
\catcode237=13 \def í{\`I}        		   
\catcode238=13 \def î{\iota}        	     
\catcode239=13 \def ï{\Psi}                     
\catcode240=13 \def ð{\times}                  
\catcode241=13 \def ñ{\Lambda}             
\catcode242=13 \def ò{\cdots}                
\catcode243=13 \def ó{\^U}			
\catcode244=13 \def ô{\`U}    	              
\catcode245=13 \def õ{\bo}                       
\catcode246=13 \def ö{\relax\ifmmode\hat\else\expandafter\^\fi}
\catcode247=13 \def÷{\relax\ifmmode\tilde\else\expandafter\~\fi}
\catcode248=13 \def ø{\ll}                         
\catcode249=13 \def ù{\gg}                       
\catcode250=13 \def ú{\eta}                      
\catcode251=13 \def û{\kappa}                  
\catcode252=13 \def ü{\half}     		 
\catcode253=13 \def ý{\Gamma} 		
\catcode254=13 \def þ{\Xi}   			
\catcode255=13 \def ÿ{\relax\ifmmode{}^{\dagger}{}\else\dag\fi}


\def\ital#1Õ{{\it#1\/}}	     
\def\un#1{\relax\ifmmode\underline#1\else $\underline{\hbox{#1}}$
	\relax\fi}

\def\roonoo#1#2#3{\vbox{\ialign{##\crcr
	\hfil{$#3{#1}$}\hfil\crcr\noalign{\kern1pt\nointerlineskip}
	$#3{#2}$\crcr}}}

\def\tdt#1{\oon{\hbox{\bf .\kern-1pt.\kern-1pt.}}#1}   
\def\({\eqno(}



\def\õ#1{
	\screwcount\num
	\num=1
	\screwdimen\downsy
	\downsy=-1.5ex
	\mkern-3.5mu
	õ
	\loop
	\ifnum\num<#1
	\llap{\raise\num\downsy\hbox{$õ$}}
	\advance\num by1
	\repeat}
\def\upõ#1#2{\screwcount\numup
	\numup=#1
	\advance\numup by-1
	\screwdimen\upsy
	\upsy=.75ex
	\mkern3.5mu
	\raise\numup\upsy\hbox{$#2$}}



\newcount\marknumber	\marknumber=1
\newcount\countdp \newcount\countwd \newcount\countht 

%
%
\ifx\pdfoutput\undefined
\def\rgboo#1{}
\input epsf

\def\postscript#1{\special{" #1}}		
\postscript{
	/bd {bind def} bind def
	/fsd {findfont exch scalefont def} bd
	/sms {setfont moveto show} bd
	/ms {moveto show} bd
	/pdfmark where		
	{pop} {userdict /pdfmark /cleartomark load put} ifelse
	[ /PageMode /UseOutlines		
	/DOCVIEW pdfmark}
\def\bookmark#1#2{\postscript{		
	[ /Dest /MyDest\the\marknumber /View [ /XYZ null null null ] /DEST pdfmark
	[ /Title (#2) /Count #1 /Dest /MyDest\the\marknumber /OUT pdfmark}%
	\advance\marknumber by1}
\def\pdfklink#1#2{%
	\hskip-.25em\setbox0=\hbox{#1}%
		\countdp=\dp0 \countwd=\wd0 \countht=\ht0%
		\divide\countdp by65536 \divide\countwd by65536%
			\divide\countht by65536%
		\advance\countdp by1 \advance\countwd by1%
			\advance\countht by1%
		\def\linkdp{\the\countdp} \def\linkwd{\the\countwd}%
			\def\linkht{\the\countht}%
	\postscript{
		[ /Rect [ -1.5 -\linkdp.0 0\linkwd.0 0\linkht.5 ] 
		/Border [ 0 0 0 ]
		/Action << /Subtype /URI /URI (#2) >>
		/Subtype /Link
		/ANN pdfmark}{\rgb{1 0 0}{#1}}}
%
%
\else
\def\rgboo#1{\pdfliteral{#1 rg #1 RG}}

\pdfcatalog{/PageMode /UseOutlines}		
\def\bookmark#1#2{
	\pdfdest num \marknumber xyz
	\pdfoutline goto num \marknumber count #1 {#2}
	\advance\marknumber by1}
\def\pdfklink#1#2{%
	\noindent\pdfstartlink user
		{/Subtype /Link
		/Border [ 0 0 0 ]
		/A << /S /URI /URI (#2) >>}{\rgb{1 0 0}{#1}}%
	\pdfendlink}
\fi

\def\rgbo#1#2{\rgboo{#1}#2\rgboo{0 0 0}}
\def\rgb#1#2{\mark{#1}\rgbo{#1}{#2}\mark{0 0 0}}
\def\pdflink#1{\pdfklink{#1}{#1}}
\def\xxxlink#1{\pdfklink{[arXiv:#1]}{http://arXiv.org/abs/#1}}

\catcode`@=11

\def\wlog#1{}	


\def\makeheadline{\vbox to\z@{\vskip-36.5\p@
	\line{\vbox to8.5\p@{}\the\headline%
	\ifnum\pageno=\z@\rgboo{0 0 0}\else\rgboo{\topmark}\fi%
	}\vss}\nointerlineskip}
\headline={
	\ifnum\pageno=\z@
		\hfil
	\else
		\ifnum\pageno<\z@
			\ifodd\pageno
				\tenrm\romannumeral-\pageno\hfil\lefthead\hfil
			\else
				\tenrm\hfil\righthead\hfil\romannumeral-\pageno
			\fi
		\else
			\ifodd\pageno
				\tenrm\hfil\righthead\hfil\number\pageno
			\else
				\tenrm\number\pageno\hfil\lefthead\hfil
			\fi
		\fi
	\fi}

\catcode`@=12

\def\righthead{\hfil} \def\lefthead{\hfil}
\nopagenumbers


\def\chrulefill{\rgb{1 0 0}{\hrulefill}}
\def\cdotfill{\rgb{1 0 0}{\dotfill}}
\newcount\area	\area=1
\newcount\cross	\cross=1
\def\volume#1\par{\newpage\noindent{\biggest{\rgb{1 .5 0}{#1}}}
	\par\nobreak\bigskip\medskip\area=0}
\def\chapskip{\par\ifnum\area=0\bigskip\medskip\goodbreak
	\else\newpage\fi}
\def\chapy#1{\area=1\cross=0
	\xdef\lefthead{\rgbo{1 0 .5}{#1}}\vbox{\biggerer\offinterlineskip
	\line{\chrulefill¼\hphantom{\lefthead}\chrulefill}
	\line{\chrulefill¼\lefthead\chrulefill}}\par\nobreak\medskip}
\def\chap#1\par{\chapskip\bookmark3{#1}\chapy{#1}}
\def\sectskip{\par\ifnum\cross=0\bigskip\medskip\goodbreak
	\else\newpage\fi}
\def\secty#1{\cross=1
	\xdef\righthead{\rgbo{1 0 1}{#1}}\vbox{\bigger\offinterlineskip
	\line{\cdotfill¼\hphantom{\righthead}\cdotfill}
	\line{\cdotfill¼\righthead\cdotfill}}\par\nobreak\medskip}
\def\sect#1 #2\par{\sectskip\bookmark{#1}{#2}\secty{#2}}
\def\subsectskip{\par\ifdim\lastskip<\medskipamount
	\bigskip\medskip\goodbreak\else\nobreak\fi}
\def\subsecty#1{\noindent{\sectfont{\rgbo{.5 0 1}{#1}}}\par\nobreak\medskip}
\def\subsect#1\par{\subsectskip\bookmark0{#1}\subsecty{#1}}
\long\def\x#1 #2\par{\hangindent2\parindent%
\mark{0 0 1}\rgboo{0 0 1}{\bf Exercise #1}\\#2%
\par\rgboo{0 0 0}\mark{0 0 0}}
\def\refs{\bigskip\noindent{\bf \rgbo{0 .5 1}{REFERENCES}}\par\nobreak\medskip
	\frenchspacing \parskip=0pt \refrm \baselineskip=1.23em plus 1pt
	\def\ital##1Õ{{\refit##1\/}}}
\long\def\twocolumn#1#2{\hbox to\hsize{\vtop{\hsize=2.9in#1}
	\hfil\vtop{\hsize=2.9in #2}}}


\twelvepoint
\font\bigger=cmbx12 \sca2
\font\biggerer=cmb10 \sca5
\font\biggest=cmssdc10 scaled 3250
 \sca5

 \sca3


\def Ü{\relax\ifmmode\Rightarrow\else\expandafter\subsect\fi}
\def Û{\relax\ifmmode\Leftrightarrow\else\expandafter\sect\fi}
\def Ú{\relax\ifmmode\Leftarrow\else\expandafter\chap\fi}

\def\itemize#1 {\item{\bf#1}}
\def\itemizze#1 {\itemitem{\bf#1}}
\def\itemutem{\par\indent\indent \hangindent3\parindent \textindent}
\def\itemizzze#1 {\itemutem{\bf#1}}
\def ª{\relax\ifmmode\leftrightarrow\else\itemizze\fi}
\def Á{\relax\ifmmode\gets\else\itemizzze\fi}

\def\¢{\ominus}

\def\Ä{\varphi}  \def\¿{\varpi}	\def\Ï{\vartheta}

\def ò{\relax\ifmmode\cdots\else\dotfill\fi}

\chardef\slo="1C


\def\cvrule{\rgbo{0 .5 1}{\vrule}}
\def\chrule{\rgbo{0 .5 1}{\hrule}}
\def\boxit#1{\leavevmode\thinspace\hbox{\cvrule\vtop{\vbox{\chrule%
	\vskip3pt\kern1pt\hbox{\vphantom{\bf/}\thinspace\thinspace%
	{\bf#1}\thinspace\thinspace}}\kern1pt\vskip3pt\chrule}\cvrule}%
	\thinspace}
\def\Boxit#1{\noindent\vbox{\chrule\hbox{\cvrule\kern3pt\vbox{
	\advance\hsize-7pt\vskip-\parskip\kern3pt\bf#1
	\hbox{\vrule height0pt depth\dp\strutbox width0pt}
	\kern3pt}\kern3pt\cvrule}\chrule}}




\def\today{\ifcase\month\or
 January\or February\or March\or April\or May\or June\or July\or
 August\or September\or October\or November\or December\fi
 \space\number\day, \number\year}

\parindent=20pt
\newskip\normalparskip	\normalparskip=.7\medskipamount
\parskip=\normalparskip	



\catcode`\|=\active \catcode`\<=\active \catcode`\>=\active 
\def|{\relax\ifmmode\delimiter"026A30C \else$\mathchar"026A$\fi}
\def<{\relax\ifmmode\mathchar"313C \else$\mathchar"313C$\fi}
\def>{\relax\ifmmode\mathchar"313E \else$\mathchar"313E$\fi}


%
%
%
%
%
%
%

\def\thetitle#1#2#3#4#5{
 \def\titlefont{\biggest} \font\footrm=cmr10 \font\footit=cmti10
  \twelverm
	{\hbox to\hsize{#4 \hfill YITP-SB-#3}}\par
	\vskip.8in minus.1in {\center\baselineskip=2.2\normalbaselineskip
 {\titlefont #1}\par}{\center\baselineskip=\normalbaselineskip
 \vskip.5in minus.2in #2
	\vskip1.4in minus1.2in {\twelvebf ABSTRACT}\par}
 \vskip.1in\par
 \narrower\par#5\par\unnarrower\vskip3.5in minus3.3in\eject}
\def\paper\par#1\par#2\par#3\par#4\par#5\par{
	\thetitle{#1}{#2}{#3}{#4}{#5}} 
\def\author#1#2{#1 \vskip.1in {\twelveit #2}\vskip.1in}
\def\YITP{C. N. Yang Institute for Theoretical Physics\\
	State University of New York, Stony Brook, NY 11794-3840}
\def\WS{W. Siegel\footnote{$*$}{
	\pdflink{mailto:siegel@insti.physics.sunysb.edu}\\
	\pdfklink{http://insti.physics.sunysb.edu/\~{}siegel/plan.html}
	{http://insti.physics.sunysb.edu/\noexpand~siegel/plan.html}}}


\pageno=0

\paper

\vskip-.5in
{\rgb{1 0 1}{Amplitudes for left-handed strings}}

\author\WS\YITP
\vskip-.1in

15-45

December 15, 2015

We consider a class of string-like models introduced previously where all modes are left-handed, all states are massless, T-duality is manifest, and only a finite number of orders in the string tension can appear.  These theories arise from standard string theories by a singular gauge limit and associated change in worldsheet boundary conditions.  In this paper we show how to calculate amplitudes by using the gauge parameter as an infrared regulator.  The amplitudes produce the Cachazo-He-Yuan delta-functions after some modular integration; the Mason-Skinner string-like action and amplitudes arise from the ÓzeroÕ-tension (infinite-slope) limit.  However, without the limit the amplitudes have the same problems as found in the Mason-Skinner formalism.

\pageno=2

Ü1. Introduction

Recently Hohm, Zwiebach, and the present author (henceforth cited as ``HSZ") [1] found a singular gauge for string theory that apparently restricted it to the massless sector, giving a closed and self-consistent truncation of the equations of motion for external fields that automatically terminated at 3 orders in $Œ'$.  (Only the bosonic string was treated there, although the method generalizes easily.)  This was a consequence of the fact that both left-handed and would-be-right-handed modes in that gauge depend only on $z$ and not $Ðz$.  (The approach was motivated by T-duality, but we will not discuss that property here, nor the related part of the formalism.)  Amplitudes were not treated.

Soon after, in a related but independent development, Cachazo, He, and Yuan (CHY) [2] developed string-like prescriptions for massless amplitudes:  They involved integrals over only $z$ that were fixed by $¶$ functions at the same values found by Gross and Mende (GM) [3] for the JWKB approximation of high-energy, fixed-angle string amplitudes.  

Shortly later a (super)stringy derivation was given for these massless amplitudes by Mason and Skinner (MS) [4], which was immediately generalized to the pure-spinor superstring by Berkovits [5] and perhaps to loops by Adamo, Casali, and Skinner (ACS) [6].  The MS-related string results [4-6], although applied only to closed (super)strings, involve only the $z$ coordinate in amplitude evaluation, using an HSZ-type action.  Although HSZ already used propagators depending only on $z$, MS went further and defined vertex operators with only $z$ integration.  The CHY $¶$ functions were part of the vertex operators.  The same BRST-invariant operator insertions $¶(P^2)$ used by ACS for loops can be used for trees, and produce the CHY/MS $¶(kÉP)$ upon collision with the ``standard" string vertex operators.  The MS-related results were successful only for Type II superstrings, and with less supersymmetry for closed string sectors where 3-point vertices received no $Œ'$ corrections (N=1 super Yang-Mills and N=0 scalars).

In this paper we will use an approach closer to the standard one for strings:  By treating the HSZ gauge more carefully as a singular limit, we find ``infinitesimal" dependence on $Ðz$ is required for IR regularization, where the gauge parameter is the IR regulator.  (Operator insertions, such as those used by MS/ACS, can also act as IR regularization.) Then integration over $Ðz$ with standard string vertex operators produces the CHY $¶$ functions.  The vertex operators are standard because the BRST operator is; even the string action is the standard one except for the gauge choice.  Our results can be applied to all strings, but with the same limitations as the MS approach.

Ü2. Tension dependence

The MS action differs from the standard one by choice of HSZ gauge, but also by taking the $Œ'£¥$ limit:  The standard Virasoro constraints are $X'ÉP$ and $P^2+X'^2/Œ'^2$, with $Œ'$ inserted according to the usual dimensional analysis; MS keep the first constraint (effectively $»/»§$) while replacing the second by just $P^2$, thus $Œ'£¥$.  (The relation to the standard action is not explicit in [4], since half the gauge invariance has already been fixed; but it follows from the BRST operator, which is standard except for $Œ'£¥$.
After this partial gauge fixing, the action has the same form as that of the null superstring [7].)
In amplitude calculations in the HSZ theory, the $Œ'£¥$ limit affects only the vertex operators, by taking $PàX'/Œ'£P$. 

By not taking this limit, the HSZ gauge gives massless amplitudes corresponding to those from field theory actions from all string theories.  Unlike the CHY/MS results, these have nontrivial (but finite-order) dependence on $Œ'$, except for the cases where the field theory has no $Œ'$ corrections for the 3-point vertices (Type II, heterotic Yang-Mills, or massless compactification scalars).
Specifically, the HSZ bosonic field theory action was of the form $R+Œ'R^2+Œ'^2R^3$ (kinematics preclude higher powers at 3-point), which is truncated by 1 supersymmetry to $R+Œ'R^2$ and by 2 supersymmetries to just $R$.  (When including the 2-form, there are also Lorentz Chern-Simons contributions [8].)  The MS string action is explicitly the limit $Œ'£¥$, and the corresponding limit on amplitudes has no analog in terms of a (super)gravity action, except for Type II.  Similar remarks apply to Yang-Mills, which is $F^2+Œ'F^3$, truncated to $F^2$ by 1 supersymmetry, and (massless) scalars, which are already restricted by kinematics to $Ä^3$ at 3-point.

So in principle we have
$$ \vcenter{
\halign{ \indent # & \qquad #\hfil & \qquad #\hfil \cr
usual & all $Œ'$ dependence & string field theory \cr
HSZ & perturbative in $1/Œ'$ & particle field theory \cr
MS & $Œ'£¥$ & field theory only for trivial $Œ'$ \cr}
} $$
However, so far we have been able to produce consistent $Œ'$ corrections only for 3-point amplitudes.

Ü3. Gauges

There are two ingredients to the recipe for calculating amplitudes for the HSZ theory from standard string actions:
(1) the singular gauge limit that defines HSZ, and
(2) corresponding boundary conditions for the conformal field theory, which give only massless states.

We begin with the bosonic string Lagrangian in Hamiltonian form, in an arbitrary gauge, and in a flat background,
$$ L_H = -ÀX^m P_m  +Â_0 ü\left(Œ'ú^{mn}P_m P_n +{1\over Œ'}ú_{mn}X'^m X'^n\right) +Â_1 X'^m P_m $$
$$ Â_0 = {å{-g}\over g_{11}}¼,âÂ_1 = {g_{01}\over g_{11}} $$
For now we set $Œ'=1$.  (It can easily be reinserted by appropriately scaling the spacetime metric and its inverse, or $X$ and $P$.)  We can also write the corresponding Lagrangian form
$$ L_L = -ü(»_R X)É(»_L X) = üå{-g}g^{MN}(»_M X)É(»_N X) $$
$$ »_{L,R} = {1\over å{Â_0}}\left[ »_  - (Â_1¦Â_0)»_§ \right]¼,âz_{L,R} = àü{1\over å{Â_0}}\left[ (Â_1àÂ_0)  + § \right] $$
(We have essentially used a zweibein in a convenient local Lorentz gauge.)  Here $z_{L,R}$ are conjugate to $»_{L,R}$ only in gauges where the worldsheet metric is constant, which are sufficient for our purposes.  Thus for later use it will be convenient to begin with the conformal gauge ($Â_0=1,Â_1=0$), and make the nonconformal, but linear, coordinate transformations
$$ z £ z_L¼,âÐz £ z_R $$
at a convenient stage to effectively change to an arbitrary constant-metric gauge.  ($z=ü( +§)$ becomes $ü( +i§)$ after Wick rotation.)  In the amplitudes, this is effectively a change of dummy variables; it introduces a gauge parameter into a gauge-invariant quantity as a way of defining an expansion.

There are 2 different families of gauges we can consider that are equivalent, but suggest different interpretations.  The first family interpolates between the conformal gauge and the HSZ gauge, where all modes are left handed:
$$ Â_0 = {1\over 1+º}¼,âÂ_1 = {º\over 1+º}¼;â⺠³ 0 $$
where
$$ º = \cases{ 0 & conformal gauge \cr ¥ & HSZ gauge \cr} $$
$$ L_H = \cases{ -ÀXÉP +ü(P^2+X'^2) & conformal gauge \cr -PÉлX & HSZ gauge \cr} $$
The interpolation between the 2 gauges looks simpler in Lagrangian form:
$$ L_L = -ü\left[ º(лX)^2 + (лX)(»X) \right] $$
The singularity of the coefficient in the HSZ gauge enforces $лX=0$.  (The limit of the Gaussian in the functional integral is a $¶$-functional.)
The rest of the analysis (ghosts, etc.) proceeds as usual, but is derived more conveniently by taking the usual conformal gauge results and making the above coordinate transformation.  These gauges correspond to the substitutions from conformal gauge
$$ z £ z_L = å{1+º}z¼,âÐz £ z_R = {1\over å{1+º}}(Ðz-ºz) $$
or after a global Lorentz transformation
$$ z_L = z¼,âz_R = Ðz-ºz $$

The other interesting family of gauges is
$$ Â_0 = º¼,âÂ_1 = 0¼;â⺠³ 0 $$
where $º=1$ is the conformal gauge.  (This is just an opposite scaling of $ $ and $§$.)  Then
$$ L_H = -ÀXÉP +ºü(P^2+X'^2) $$
$$ L_L = -ü\left( {1\over º}ÀX{}^2 -ºX'^2 \right) $$
The interesting singular gauges are now $º=0$ or $¥$, which enforce $ÀX=0$ or $X'=0$.  The relevant coordinate transformation is
$$ z_{L,R} = ü\left( 庠 à {1\over åº}§ \right) $$

%

Ü4. Boundary conditions

Usually $L_0-ÐL_0$ is used to constrain the left and right excitation levels to be related.  But for HSZ it also determines the $ $ development, which in turn determines the boundary conditions:  $ $ development determines the ``energy" of the operators appearing in the mode expansion, and creation operators are those that carry positive energy.  Equivalently, annihilation operators have coefficients that vanish as $ £-¥$, i.e., $z£0$.  In other words,
$$ L_0 - ÐL_0 = {»\over »§} = \left({»\over »§}\right)_L + \left({»\over »§}\right)_R $$
and both terms have the same oscillator expansion, with the ÓsameÕ sign, since in this gauge $»/»§$ and $»/» $ are identified.
In terms of R oscillators, this means $a£aÿ,aÿ£-a$ (with opposite sign to preserve commutation relations).

The net result is that left and right creation operators contribute with the same sign to $L_0-ÐL_0$.  With appropriate fixing of the normal-ordering constant, this fixes the ÓsumÕ of left and right excitation levels to be 2.  (I.e., both L and R $»/»§$'s have the same oscillator expansion as the open string's $L_0$, and get the same sign contribution.)  This restricts vertex operators to go ÓonlyÕ as $ZZ$ or $Z'$, where $Z­(PàX')/å2$, exactly those considered in [1] to describe the massless sector.  (This is easily extended to include ghost modes, which are needed to couple the dilaton.  But this string-frame dilaton is a constant in unitary gauges such as lightcone, where the physical scalar is contained in the trace/determinant of the metric.)

On shell, at the linearized level, in Lorenz gauges, the $Z'$ terms can be dropped, and $ZZ$ can be restricted to $Z_L Z_R­ü(P+X')(P-X')$, corresponding to the massless states of the usual closed bosonic string.

Similarly, there is a sign change in terms of oscillators for $L_0+ÐL_0$.  On the other hand, the zero-modes contribute with the same sign from the 2 terms, whereas they cancel for $L_0-ÐL_0$.  This was required for the absence of winding modes.

The usual conformal-gauge propagator $ÒXÊXÔ$ is
$$ ë_0 = -\ln (Ðzz+·) $$
The main difference between the conformal gauge and the HSZ gauge is the switching of the roles of $L_0+ÐL_0$ with $L_0-ÐL_0$; so effectively $ÐL_0£-ÐL_0$.  Thus we need to change the sign of the $Ðz$ part of the propagator.  (This sign has the same explanation in terms of oscillators as above.)  This can be accomplished by changing the boundary conditions, adding a ÓhomogeneousÕ solution to the propagator:
$$ ë_0 £ ë = ë_0 + 2\ln Ðz = \ln \left( {Ðz\over z +·/Ðz} \right) $$
Making the coordinate transformation relevant for the HSZ gauge,
$$ ë £ \ln \left[ {Ðz-ºz\over z +·/(Ðz-ºz)} \right] $$
Dropping an irrelevant constant, the final result is
$$ ë = \ln \left[ 1 - {1\over º}Ê{Ðz \over z + ·/(Ðz-ºz)} \right] $$
($Œ'$ can be restored by rescaling so $뾌'$.)  When we consider the HSZ limit, we have simply
$$ \lim_{º£¥} ë = - {Ðz\over º}Ê{1\over z} $$
Formally this vanishes in the limit $º£¥$, in agreement with HSZ.  But $º$ acts as an IR regulator for $Ðz$ integration for 4- and higher-point amplitudes, as we'll see below.  (It thus takes the place of the corresponding insertions of MS/ACS.)  

Any propagators for (anti)chiral fields (ghosts, physical fermions, chiral factors of vertex operators) receive similar modifications: 
$$ {1\over z} £ {1\over z}¼,â{1\over Ðz} £ {1\over º} \left( {1\over z} + {Ðz\over º}Ê{1\over z^2} \right) $$
and similarly for powers of these (but note there is an overall sign change for antichiral propagators).  As for $ë$, we have terminated the approximation at first order in $Ðz/º$.  (We'll see that, except for a trivial rescaling, $Ðz$ and $º$ always come in this combination.)
These results also apply to (the HSZ analog of) lightcone gauge, which is a special case of conformal gauge.
The net effect is that antichiral operators have effectively become chiral (except for regularization), in line with the HSZ interpretation that the entire theory is left-handed.  Another interpretation, represented by our other family of gauges, is that all operators are functions of only $ $, so the string theory has been reduced to a particle theory.

At tree level, only ghost zero-modes are relevant, so their propagators are not needed for calculating amplitudes.  The zero-modes saturate the vacuum in the usual way, as can be derived by expansion and ordinary integration over them.  

To see nontrivial $ÒXÊXÔ$ contributions, we consider the contribution of just the usual $e^{ikÉX}$'s to an amplitude:
$$ Þ_i Òe^{ik_iÉX(z_i)}Ô = e^{S_0} $$
$$ S_0 ¼=¼ {1\over 2º} Ý_{i,j} k_iÉk_j {Ðz_{ij}\over z_{ij}} ¼=¼ {1\over º} Ý_i Ðz_i Ý_j {k_iÉk_j \over z_{ij}} $$
Integrating over $Ðz_i$,
$$ Ç_{-¥}^¥ d^{n-3} Ðz_iÊe^{S_0} ¼¾¼ º^{n-3} Þ_i ¶ \left( Ý_j {k_iÉk_j \over z_{ij}} \right) $$
These are just the CHY factors that enforce the GM conditions.  While the MS approach effectively introduced such factors explicitly through the vertex operators, we find it more convenient (and more similar to conventional string theory) to obtain them after integrating propagators over the usual moduli.  
Note that the limit $º£¥$ must be taken before any $Ðz$ integrals are performed, since the exact amplitude is $º$ independent; the limit gives the first term in a $1/º$ expansion.  (This expansion is equivalently an expansion in $Ðz$.  The $º^{-n}$ from the R vertex operators and the $º^3$ from the R ghosts cancel the $º$ dependence, due to conformal invariance.)

Ü5. Tree amplitudes

So the net modification to propagators is
$$ ÒXÊXÔ £ - {Ðz\over º}Ê{1\over z}¼;ââ{1\over z} £ {1\over z}¼,â{1\over z^2} £ {1\over z^2} $$
$$ {1\over Ðz} £ {1\over º} \left( {1\over z} + {Ðz\over º}Ê{1\over z^2} \right)
	¼,â{1\over Ðz^2} £ -{1\over º^2} \left( {1\over z^2} + 2Ê{Ðz\over º}Ê{1\over z^3} \right) $$
The rules for evaluating amplitudes in the limit $º£¥$ are then:\\
(1) Evaluate the conformal-gauge amplitude, except for performing $z$-$Ðz$ integration.\\
(2) Substitute the above propagators.\\
(3) $Çd^{n-3}Ðz$ to get (modified) CHY $¶$-functions.\\
(4) $Çd^{n-3}zʶ^{n-3}$ to fix the values of all the $z$'s in terms of momentum invariants.

As in standard string theory (see, e.g., [9]), such amplitudes can be evaluated by exponentiating the $»_{L,R}X$ factors of the vertex operators:
$$ V = {1\over º}Çd^2½¼e^{ikÉX + ½·É»_L X + ºÐ½Ð·É»_R X} $$
for anticommuting variables $½$.  (We have included canceling factors of $º$ in the definition in anticipation of that from $1/z_R$.)

The polarizations have been factorized as
$$ ·^{mn} = ·^m з^n $$
Technically they are a sum of such terms, but the vectors can be reassembled into tensors at the end of the calculation.
To make the exponent bosonic, rather then introduce more fermionic integration variables, we have taken the vector $·$'s as anticommuting:  The tensor ones are then still commuting.  
(The resulting signs obtained in reassembling the tensors conveniently take the place of ordering signs that would be required if the vectors were taken as bosonic, making the corresponding part of the exponent fermionic.  It's interesting to note that in T-duality-manifest form, where L and R indices are combined into an SO(D,D) index $M$, $·^{MN}=·^M ·^N$ is antisymmetric by Fermi symmetry, which is consistent with the fact that the doubled metric or vielbein is orthogonal, and thus linearizes as antisymmetric.)

The amplitude is then naively, after the $Ðz$ integrals have been evaluated as above,
$$ A ¼¾¼ (z_{12}z_{13}z_{23})^2 Çd^{n-3}z Þ_{i=4}^n ¶ \left( Ý_j E_{ij} \right) 
	\left( Çd^n ½_LÊe^{S_L} \right) \left( Çd^n ½_RÊe^{S_R} \right)  $$
$$ E_{ij} = {k_iÉk_j \over z_{ij}} + ... $$
$$ S ¼=¼ Ý_{i.j} \left( à üÊ {½_i ½_j ·_iÉ·_j \over z_{ij}^2} + i å{Œ'}¼{ k_iɽ_j ·_j \over z_{ij}} \right) $$
where ``$S$" stands for $S_L$ or $S_R$ with correspondingly $½·$ or $нз$, ``$à$" is + for L and $-$ for R, and we have restored $Œ'$ dependence.  We can neglect $i=j$ terms, or apply
$$ k_iÉk_i = k_iÉ·_i = ·_iÉ·_i = 0 $$
(Note that 2 $kÉ·$ terms are required to replace an $·É·$ term, so orders in $Œ'$ are always integer.  Since the $·$'s are fermionic, $·_iÉ·_i=0$ is automatic.)

We can also replace the L and/or R vertex factors with massless compactification scalar vertex factors (as in the bosonic chirality of the heterotic string):  The currents $(PàX')/å2$ are then replaced with currents for the gauge group.  Writing the gauge group as a subgroup of U(N), we can then write the terms in the exponential of the vertex operator as
$$ ½·É»_{L,R}X £ ½ÐÆ^a ·_a{}^b Æ_b $$
(L and/or R) in terms of fermionic fields $Æ,ÐÆ$.  Since the coupling is now quadratic rather than linear, the functional integral will give a determinant instead of an exponential,
$$ \det \left( ¶_i^j ¶_a^b + {1\over z_{ij}}ʽ_j ·_{ja}{}^b \right) $$
(for a matrix $M_{ia}{}^{jb}$).  This can conveniently be rewritten as usual as $det =$ $exp$ $tr$ $ln$, and the group trace of $·$'s can be rewritten as a Chan-Paton trace of adjoint representation matrices (structure constants).

The calculation of the 3-point function is identical to the usual string calculation, since there is no $z$ nor $Ðz$ integration (but still $½$), and according to massless kinematics all $k_iÉk_j=0$.  The vertices are those arising from (the T-duality completion of) a gravity Lagrangian of the form $R+R^2+R^3$ (where $R^2$ is the Gauss-Bonnet combination so the propagator is still $1/p^2$).  $k$ contributions from $S$ are necessary because an odd number (3) of $½_L$'s and of $½_R$'s are needed.  These corrections resemble those of a ``heterotic"-like theory [7], as should have been expected from the left-handed starting point.

Generalization to superstrings (in either RNS or manifestly supersymmetric formulations) is straightforward, following the above steps for modifying the gauge choices and propagators.

But at 4-point and higher, the same problems are found as in MS, except for the cases they explicitly evaluated, where they reproduced the CHY amplitudes.  As mentioned above, these are exactly the cases where the 3-point receives no $Œ'$ corrections.

Ü6. Loops

Similar methods can be applied to loops.  (At short distances, the modifications to the propagator should be the same as for trees.)  Generically, the conformal gauge will give an exponent of the form
$$ E = Ý_{i,j}k_iÉk_jG_{ij}¼,âG_{ij} = f(z_i,z_j) + Ðf(Ðz_i,Ðz_j) + \hbox{0-mode terms} $$
We then add a homogeneous term that changes the sign of the $Ðz$ term, and substitute
$$ z £ z¼,âÐz £ z - {1\over º}Ðz $$
In the limit $º£¥$, this has the net effect
$$ E £ {1\over º} Ý_i Ðz_i {»\over »z_i} E $$
So integrating over the $Ðz$'s gives
$$ Þ_i ¶\left( Ý_j k_iÉk_j {»\over »z_i} G_{ij} \right) $$
enforcing the same conditions as from a JWKB approximation with respect to $z$ integration.  These $¶$'s fix the positions of the vertex operators, leaving integration over only the moduli that specify the conformal geometry of the worldsheet.

Ü7. Conclusions

Our modification of string theory involves only the use of a singular gauge and especially a change in worldsheet boundary conditions.  (If we were to change only the gauge but not change the boundary conditions, we would obtain the amplitudes of the ``new dual model" of [10].)

So far we have managed only to give a derivation of the HSZ and MS results from ordinary string theory, explaining the truncation to massless states (and the chiral nature of the results) by a change in worldsheet boundary conditions.  In the future we hope we further analyze the method, to allow a generalization to other amplitudes or include $Œ'$ corrections.

ÜAcknowledgments

This work was supported in part by National Science Foundation Grant No.¼PHY-1316617.  I thank Nathan Berkovits for pointing out fatal errors in an early version of this work.

\refs

£1 O. Hohm, W. Siegel and B. Zwiebach,
  ``Doubled $Œ'$-Geometry,''
  JHEP {\bf 1402} (2014) 065
  \xxxlink{1306.2970} [hep-th].

£2 F. Cachazo, S. He and E.Y. Yuan,
  ``Scattering of Massless Particles in Arbitrary Dimension,''
  Phys.\ Rev.\ Lett.\  {\bf 113} (2014) 17,  171601
  \xxxlink{1307.2199} [hep-th];\\
  ``Scattering of Massless Particles: Scalars, Gluons and Gravitons,''
  JHEP {\bf 1407} (2014) 033
  \xxxlink{1309.0885} [hep-th].

£3 D.J. Gross and P.F. Mende,
  ``The High-Energy Behavior of String Scattering Amplitudes,''
  Phys.\ Lett.\ B {\bf 197} (1987) 129;\\
  ``String Theory Beyond the Planck Scale,''
  Nucl.\ Phys.\ B {\bf 303} (1988) 407.

£4 L. Mason and D. Skinner,
  ``Ambitwistor strings and the scattering equations,''\\
  JHEP {\bf 1407} (2014) 048
  \xxxlink{1311.2564} [hep-th].

£5 N. Berkovits,
  ``Infinite Tension Limit of the Pure Spinor Superstring,''
  JHEP {\bf 1403} (2014) 017
  \xxxlink{1311.4156} [hep-th].

£6  T. Adamo, E. Casali and D. Skinner,
  ``Ambitwistor strings and the scattering equations at one loop,''
   JHEP {\bf 1404} (2014) 104
 \xxxlink{1312.3828} [hep-th].

£7 I. Bandos,
  ``Twistor/ambitwistor strings and null-superstrings in spacetime of D=4, 10 and 11 dimensions,''
  JHEP {\bf 1409} (2014) 086
  \xxxlink{1404.1299} [hep-th].

£8 O. Hohm and B. Zwiebach,
  ``Green-Schwarz mechanism and $\alpha'$-deformed Courant brackets,''
   JHEP {\bf 1501} (2015) 012
 \xxxlink{1407.0708} [hep-th];\\
  ``Double Field Theory at Order $\alpha'$,''
  JHEP {\bf 1411} (2014) 075
  \xxxlink{1407.3803} [hep-th].

£9 M.B. Green, J.H. Schwarz and E. Witten,
  ``Superstring Theory, Vol. 1: Introduction''
  (Cambridge University, 1987) pp. 374, 398.

£10 N.E.J. Bjerrum-Bohr, P.H. Damgaard, P. Tourkine and P. Vanhove,
  ``Scattering Equations and String Theory Amplitudes,''
  Phys.\ Rev.\ D {\bf 90} (2014) 10,  106002\\
  \xxxlink{1403.4553} [hep-th].

\bye